\documentclass[aps,prl,twocolumn,showpacs,superscriptaddress]{revtex4}

\usepackage{graphicx} % Include figure files
\usepackage{longtable}

\begin{document}

\title
{\emph{Ab-initio} electron transport calculations of carbon based
string structures}

\author{S. Tongay}
\affiliation{Department of Physics, Bilkent University, 06800
Ankara, Turkey}
\author{R.T. Senger}
\affiliation{Department of Physics, Bilkent University, 06800
Ankara, Turkey} \affiliation{T{\"U}B{\.I}TAK - UEKAE, P.K.74,
41470 Gebze, Kocaeli, Turkey}
\author{S. Dag}
\affiliation{Department of Physics, Bilkent University, 06800
Ankara, Turkey}
\author{S. Ciraci} \email{ciraci@fen.bilkent.edu.tr}
\affiliation{Department of Physics, Bilkent University, 06800
Ankara, Turkey}

\date{\today}

\begin{abstract}
First-principles calculations show that monatomic strings of
carbon have high cohesive energy and axial strength, and exhibit
stability even at high temperatures. Due to their flexibility and
reactivity, carbon chains are suitable for structural and chemical
functionalizations; they form also stable ring, helix, grid and
network structures. Analysis of electronic conductance of various
infinite, finite and doped string structures reveal fundamental
and technologically  interesting features.  Changes in doping and
geometry give rise to dramatic variations in conductance. In
even-numbered linear chains strain induces substantial decrease of
conductance. The double covalent bonding of carbon atoms underlies
their unusual chemical, mechanical and transport properties.
\end{abstract}

%
%%73.22.-f   Electronic structure of nanoscale materials:  clusters,
%%            nanoparticles, nanotubes, and nanocrystals
%%61.48.+c   Fullerenes and fullerene-related materials
%%73.20.Hb   Impurity and defect levels; energy states of adsorbed species
%%71.30.+h   Metal-insulator transitions and other electronic transitions
%%71.20.Tx   Fullerenes and related materials; (Band str.)
%%71.15.Nc   Total energy and cohesive energy calculations
%

\maketitle

 While the research on molecular electronics is in
progress since 1960s, the problem of interconnecting those
molecular devices  has remained as a real challenge. Recently,
linear gold chains suspended between two gold electrodes have been
produced \cite{ohnishi98}. Another ultimate one-dimensional
nanowire, namely monatomic carbon linear chain (C-LC) has been
observed at the center of multi-wall carbon nanotubes
\cite{zhao03}. Some carbon chain structures have been subjects of
earlier theoretical studies
\cite{karphen,jones,bylaska,saito99,torelli00,abdur02}.

This Letter predicts new stable structures of carbon-based strings
and their unusual electronic transport properties. Carbon strings
having impressive mechanical and electronic properties can form
helix structures through bending, and also networks or 2D grids
through T and cross-bar junctions. Interesting and potentially
useful conductance variations can be achieved in these structures.
In spite of the fact that the parent diamond crystal is a good
insulator, an ideal C-LC  is a better conductor than gold chain.
Double-bond formation between adjacent C atoms underlies all these
unusual properties of the carbon chain.

We carried out total energy and electronic structure calculations
using first-principles pseudopotential plane wave method
\cite{vasp} within density functional theory (DFT) and supercell
geometries. Generalized gradient approximation (GGA) and local
density approximation (LDA) have been used, and all the atomic
positions and lattice parameters of string structures have been
optimized. The analysis of quantum ballistic conductance has been
performed using {\sc transiesta-c}, a recently developed
\emph{ab-initio} transport software based on localized basis sets,
DFT and non-equilibrium Green's function (NEGF) formalism
\cite{trans}.

Among a large number of nanowire structures of carbon including
planar zigzag, dumbbell, and triangular structures, C-LC (known as
cumulene) is found to be the only stable structure \cite{struct}.
Upon relaxation, all other  structures are transformed to C-LC
with a bond length of $c=1.27$ \AA.  The C-LC is energetically
very favorable as reflected by the cohesive energy of as large as
8.6 eV/atom (that is close to the calculated GGA cohesive energy
in diamond structure, 9.5 eV/atom).

\begin{figure}
\includegraphics[scale=0.36]{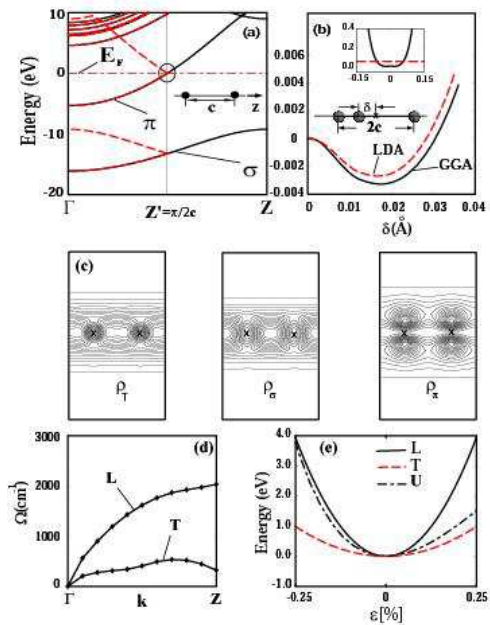}
\caption{(a) Energy band structure of C-LC. Zone-folded bands upon
doubling the cell size from $c$ to $2c$ are shown by dashed lines.
(b) Variation of LDA and GGA total energies per unit cell of C-LC
in the presence of Peierls distortion. The inset shows the
zero-point oscillation energy of a C atom in the effective
potential due to Peierls distortion. (c) Contour plots of total,
$\sigma$-band, and $\pi$-band charge-densities in the plane
passing through C-LC. (d) Longitudinal (L) and doubly degenerate
transversal (T) branches of phonons. (e) Variation of total energy
with the lateral (solid line), transversal (dashed line)
displacements of a single C atom in a supercell of four atoms.
Dot-dashed line (U) shows the variation of the total energy with
uniform extension or compression of $c$ from equilibrium. }
\label{fig:1}
\end{figure}

Structural and electronic properties  of C-LC are essential for
studying its transport characteristics (see Fig. \ref{fig:1}). The
$\sigma$-band below the Fermi level, $E_F$, is composed from $2s$
and $2p_z$ atomic orbitals. The doubly degenerate $\pi$-band
formed from the bonding combinations of $2p_x$ and $2p_y$ atomic
orbitals \cite{karphen}, crosses $E_F$ at $k=\pi/2c$. Therefore
the C-LC structure is vulnerable to Peierls distortion which may
induce a metal-insulator transition. The existence and degree of
such a distortion has been a subject of debate \cite{abdur02}. We
calculated that the maximum energy gain  (per atom) through the
dimerization of the chain is just 2.7 (3.2) meV using LDA (GGA) at
bond alternation parameter $\delta=0.017$\AA. However, treating
the GGA-calculated energy-change profile as an effective potential
for the dynamics of a C atom in the chain, the zero-point energy
of the atom in the double-well potential turns out to be much
higher (51 meV) than the Peierls distorted ground state energy
(see Fig. \ref{fig:1}(b)). As a result, the small distortion
effect is insignificant and cannot be observed even at T=0 K as
calculated with GGA and LDA energies \cite{distortion}.

In addition to the $\sigma$-bond derived from the states of the
$\sigma$-band, half filled $\pi$-band states form a $\pi$-bond as
shown in Fig. \ref{fig:1}(c). This double-bond structure is
responsible for the high stability and metallicity of ideal C-LC.
Calculated longitudinal vibrational modes have higher frequencies
than the doubly degenerate transverse modes as shown in Fig.
\ref{fig:1}(d). This indicates that the chain is stiff along its
axis but flexible in the transverse directions. Accordingly, C-LC
is suitable for deformations to form other string structures. The
elastic response of the C-LC to different deformations is
summarized in Fig. \ref{fig:1}(e). As a measure of the elastic
stiffness of the C-LC the second derivative of the strain energy
per atom with respect to  the axial strain,
$d^{2}E/d\epsilon^{2}$, is calculated to be 119 eV. This value
corresponds to a very high axial strength for the C-LC when
compared to the corresponding values for carbon nanotubes which
are in the range 52-60 eV \cite{portal99}.

\begin{figure}
\includegraphics[scale=0.36]{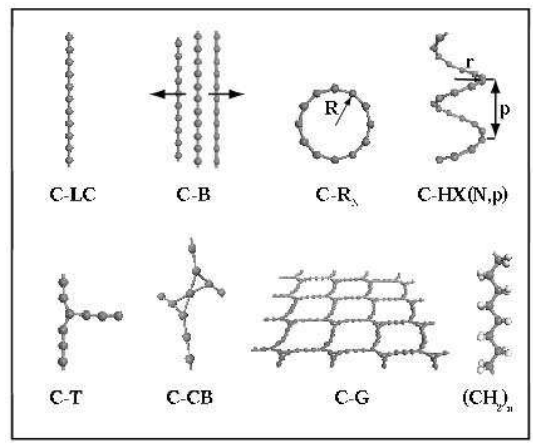}
\caption{Carbon-chain-based nanostructures: Carbon linear chain
(C-LC); bundle of several linear chains (C-B) with a repulsive
inter-chain interaction; ring structure with $N$ atoms (C-R$_N$);
helix of $N$ atoms with pitch length $p$ (C-HX${(N,p)}$);
T-junction of two chains (C-T); cross-bar structure of two chains
(C-CB); planar grid structure (C-G); and C-LC functionalized with
H-atoms ((CH$_{2}$)$_{n}$).} \label{fig:str}
\end{figure}

Even more interesting are structural and chemical
functionalizations of carbon strings as illustrated in Fig.
\ref{fig:str}. Carbon rings, C-R$_N$, containing $N$ atoms have a
number of interesting features, and is closely related with the
helix structure \cite{torelli00}. The helix structure by itself,
is denoted as C-HX$(N,p)$, and can be generated from C-LC with
different radii and pitch lengths $p$ (in \AA). Here $N$ is the
number of C atoms in one pitch of the helix. The binding energies
of helices are slightly smaller than that of the C-LC due to
strain energy implemented by the curvature, such as for $p=10$
\AA\, fixed, $E_b$ is 8.3 and 8.5 eV for $N=10$ and 16,
respectively. Finite helix structure are to be stabilized by
fixing from both ends against a weak tendency of transforming to a
LC. All C-HX structures we studied here have the bond alternation
no matter what the value of $N$ is, and are stable even at
$T=1200$ K. We believe that C helix structure may be relevant to
study chiral currents and to fabricate nanosolenoids. Further to
circular ring and helix structures, the C-LC can branch off to
form T and cross junctions. These junctions can form, since an
additional C atom can easily be attached to any carbon atom of
C-LC with a binding energy of 8 eV. For example, a T-junction is
produced if a perpendicular chain develops from that adatom. At
the junction the double bond structure changes into planar
$sp^2$-like bonding. In the case of cross-junction, two carbon
atoms are attached to a single atom of C-LC. At the junction a 3D
$sp^3$-like bonding develops as shown in Fig. \ref{fig:str}. These
chains can further branch off to produce a grid of ($N \times M$)
atomic rectangular cells \cite{cell}, or a network structure. In
principle, these grids or networks can be functionalized to
generate 2D and 3D artificial crystals. Since the interconnects
between the nodes are conducting, one can expect that carbon chain
networks or grids can be useful for  the integration in the
molecular electronics.

\begin{figure}
\includegraphics[scale=0.4]{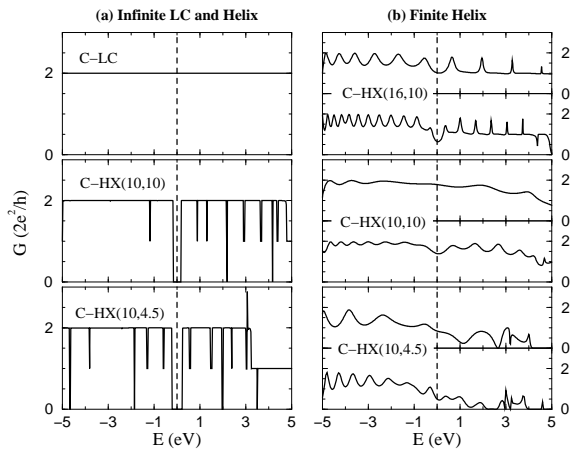}
\caption{Calculated conductance of various carbon chain structures
in units of the conductance quantum ($G_0=2e^2/h$). (a) Infinite
structures of undistorted C-LC and two helices C-HX${(N,p)}$ with
different pitch lengths. (b) Finite-sized helices in between two
C-LC electrodes. Each panel is divided into two; the top parts are
for a single period of the helix, and the bottom parts are for two
periods. The Fermi levels are set to zero in all systems.}
\label{fig:con}
\end{figure}

Carbon chains are reactive owing to their double bond structure.
Oxygen atom can be chemisorbed to a single carbon atom (specified
as top site) or form a bridge bond between two carbon atoms.
Similarly, a single hydrogen atom can easily be adsorbed to a
chain atom with $E_b=3.6$ eV; it is larger than the energy
associated with the chemisorption of H on single-wall carbon
nanotubes \cite{gulseren01}. The (CH$_2$)$_n$ structure in
Fig.~\ref{fig:str} is a wide band-gap semiconductor.

Having revealed the atomic and electronic structures we now turn
to the prime objective of our study and examine how the
conductance of these infinite and finite C-based string structures
vary with geometry, elastic deformation and doping. As usual, we
have used an electrode-device-electrode geometry for conductance
calculations. The conductance of the device has been calculated as
$G(E)=(2e^2/h)\textrm{Tr}(\Gamma_1 G^r \Gamma_2 G^a)$
($G^{\{r,a\}}$: retarded and advanced Green's functions,
$\Gamma_{1,2}$: coupling functions to the electrodes) after
iteratively solving NEGF and DFT equations of the system
\cite{trans}. In order to match the device potential and the
surface potential of the semi-infinite electrodes, the device
regions are defined to contain some portion of the electrodes.
Here, we choose to use C-LC as the metallic electrodes. The atoms
in the device region are subjected to geometry optimizations in
supercells, keeping the electrode-atom positions fixed.

Earlier, conductance properties of short C chains have been
reported. Using jellium electrodes it has been shown that the
conductance of C atomic chains varies in an oscillatory manner
with the number of atoms in the chain \cite{lang}. Larade \emph{et
al} \cite{hatem} predicted negative differential resistance for C
chains coupled with Al electrodes. They also point out
quantitative differences in odd- and even-numbered C chains.
\cite{hatem}. We investigate effects of strain, impurities and
adatoms on conductance of C strings. Results of \emph{ab-initio}
electron transport calculations of C-LC, infinite and finite
segments of C-HX with varying $N$ and $p$ are summarized in Fig.
\ref{fig:con}. The equilibrium conductance of ideal, undistorted
C-LC is $2G_0$ due to the doubly degenerate $\pi$-band crossing
the Fermi level. In case of bond alternation due to Peierls
distortion there opens a gap around $E_F$. However, according to
the present GGA results the weak bond alternation should not be
observable due to zero-point oscillations of the atoms. We also
note that any defect (such as displacement of atoms from their
equilibrium positions, etc.) in C-LC should lead to the
localization of current transporting state, which is characterized
by the localization length $\xi$. For a strictly 1D wire like
C-LC, $\xi$ is proportional to the elastic mean free path $\ell$
($\ell$ itself depends on the type of defect and temperature). If
the length $L$ of the C-LC is larger than $\xi$, the resistance of
the chain becomes $R \gtrsim \frac{h}{4e^2} +
\frac{h}{e^2}e^{L/\xi}$ \cite{altschuler}. On the other hand the
opening of a conductance gap around $E_F$ is related to the
dimerization of carbon atoms, and that effect is seen to be
reflected in the conductance plots of infinite carbon helix
structures. The bending of the carbon chain into a helix also
introduces some narrow conductance drops of integer number, which
changes with the pitch length and radius of the helix. In case of
finite-length solenoids, we considered single and double periods
of helices placed in between two semi-infinite C-LC electrodes.
The dimerization of atoms persists for finite-length helices. In
Fig. \ref{fig:con}(b) the set of curves presented for the
conductance of finite-sized helices all display oscillatory
behavior; the frequency of which depends on the radius and pitch
length of the helix. Each peak corresponds to a molecular orbital
level of the device (helix) aligned  with $E_{F}$ of the
electrodes (C-LC). As the axial size of the helix is doubled the
number of the peaks in the conductance pattern is also doubled,
pointing to the limiting case of infinite helix structure.

\begin{figure}
\includegraphics[scale=0.4]{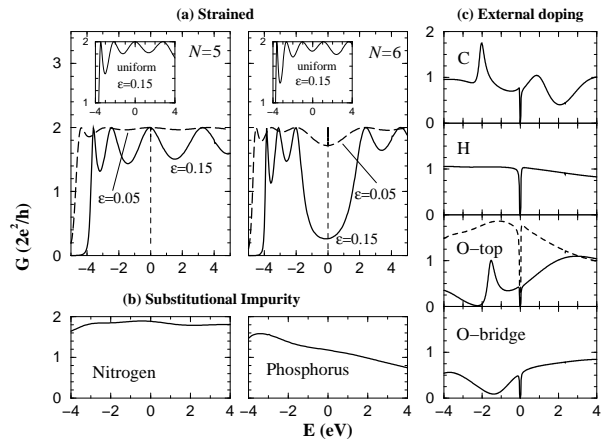}
\caption{Conductance variations of finite C-LCs modified by; (a)
strain under tension, (b) substitutional impurity atoms of N and
P, (c) chemical doping with single C or H atom; single O atom at
top site (dashed curve is obtained by deleting the adsorbed O but
by maintaining the local distortion in C-LC caused by O); single O
atom adsorbed at the bridge site. The Fermi levels are set to zero
in all systems.} \label{fig:con2}
\end{figure}

How the ballistic electron transport depends on strain,
substitutional impurity and external doping of C-LC is another
interesting aspect of our study. When an axial tension $F$ is
applied to an $N$-atom C-LC the chain elongates (the total length
scales by $1+\epsilon$, where $\epsilon$ is the strain) resulting
with non-uniform bond lengths. For example, for $F=7.78$ nN a
6-atom chain attains a strain $\epsilon=0.15$ and a sequence of
equilibrium bond lengths; 1.56, 1.31, 1.59, 1.30, 1.59, 1.31, 1.56
\AA\, when relaxed in between two C-LC electrodes. Resulting
conductance spectrum of the strained chains displays oscillatory
behavior with $E$. Fig. \ref{fig:con2}(a) presents $G(E)$ of a
C-LC device including $N=5$ or $N=6$ atoms subjected to different
strains. The character of oscillations is dramatically different
depending on $N$ being odd or even (We have confirmed that chains
with $N=4$ and 7 comply with this trend). If $N$ is odd, the chain
always has a peak conductance value of 2$G_0$ at the Fermi level,
whereas for even $N$, $G(E_F)$ corresponds to a dip of
oscillations which drops considerably with increasing $\epsilon$.
This salient feature is related to the alignment of the molecular
energy levels of the device region with the electrochemical
potential of the electrodes. It turns out that $E_F$ coincides
with the HOMO (highest occupied molecular orbital) for
odd-numbered chains, and with the center of HOMO - LUMO (lowest
unoccupied molecular orbital) gap for even-numbered ones. When
geometrical optimization of the chain is omitted and the strain is
assumed to be uniform the conductance oscillations are still
present but their amplitudes are smaller as shown in the insets of
Fig. \ref{fig:con2}(a). The observation that even-numbered
strained C chains have lower conductance values is compatible with
the findings of Lang and Avouris for unstrained chains
\cite{lang}. They report $G\cong1.8G_0$ for C chains with $N=5$
and 7 in between jellium electrodes. In our model, use of C-LC
electrodes naturally corresponds to a ``better contact" which
leads to $G=2G_0$. With our choice of electrodes, unlike the
results in Ref. \onlinecite{lang}, the difference of $N$ being odd
or even in conductance of \emph{unstrained} C chains does not
manifest itself, because for all $N$ the configuration is the
same. Therefore, the odd -- even $N$ disparity we find in
conductance of C chains is a pure strain-induced effect.

$G(E)$ is also affected by substitutional impurities as shown in
Fig. \ref{fig:con2}(b). We considered 7-atom C-LCs containing a
substitutional impurity atom of N or P, at the center. While the
decrease in conductance is minute for N-impurity, P causes a
substantial decrease. Conductance variations $G(E)$ for C-LC
externally doped by a single (C, H, and O) atom are illustrated in
Fig. \ref{fig:con2}(c). The adatom deforms the C-LC by changing
the type and the geometry of the local bonding at the adsorption
site. For all cases, we see that $G(E)$ dips to zero at $E_F$. The
dip is dominantly due to the deformation of the chain since it
persists even if the adatom is removed but the distortion of the
chain is kept the same. The $G(E)$ curve reflects the local
electronic structure at the adatom site, and is a fingerprint for
the type of the adatom, as well as the adsorption geometry.

In summary, we showed that carbon atoms can form stable string
structures with impressive physical properties. C-LC is flexible
but have very high stiffness along the axial direction.
Accordingly, carbon strings are suited for forming various  stable
nanostructures; they can be easily doped and functionalized by
chemisorption of adatoms and molecules. Calculated conductance is
found to be sensitive to deformation geometry and doping of the
chain structure. Strained C chains show an odd-even $N$ disparity.
As an interconnect carbon atomic string structures can be a
potential alternative to gold chain as well as to metallic carbon
nanotubes. We believe that with their novel electronic,
mechanical, and transport properties, carbon strings deserve
further theoretical and experimental research to be a potential
material for nanoscience and nanotechnology.


\begin{thebibliography}{99}


\bibitem{ohnishi98} H. Ohnishi \emph{et al}, Nature (London) \textbf{395}, 780 (1998).

\bibitem{zhao03} X. Zhao \emph{et al}, Phys. Rev. Lett. \textbf{90}, 187401 (2003).
\bibitem{karphen}A. Karphen, J. Phys. C: Solid State Phys. \textbf{12}, 3227 (1979).
\bibitem{jones} R. O. Jones \emph{et al}, Phys. Rev. Lett. \textbf{79}, 443 (1997).
\bibitem{bylaska} E. J. Bylaska \emph{et al}, Phys. Rev. B \textbf{58}, R7488 (1998).
\bibitem{saito99} M. Saito \emph{et al}, Phys. Rev. B \textbf{60}, 8939 (1999).
\bibitem{torelli00} T. Torelli \emph{et al}, Phys. Rev. Lett. \textbf{85}, 1702 (2000).
\bibitem{abdur02} A. Abdurahman \emph{et al}, Phys. Rev. B \textbf{65}, 115106 (2002), and references therein.

\bibitem{vasp} Numerical calculations have been performed by using {\sc vasp} software package.
G. Kresse and J. Hafner, Phys. Rev. B \textbf{47}, 558 (1993);
G. Kresse and J. Furthm\"{u}ller,  \emph{ibid} \textbf{54}, 11169 (1996).
51 special \textbf{k}-points with $|\textbf{k+G}| \cong 400$ eV have been used for
the C-LC; for other structures this number is scaled. We performed an extensive analysis
of stability by reoptimizing the structures after displacing individual atoms from
their original optimized configuration, by calculating phonon spectrum using {\sc ab-init},
and also by carrying out high temperature molecular dynamics calculations ($800$ K $\leq T \leq 1200$ K).

\bibitem{trans} The methodology of the {\sc transiesta-c} software is described in:
 M. Brandbyge \emph{et al}, Phys. Rev. B \textbf{65}, 165401 (2002).
 The software was  provided by Atomistix corp.

\bibitem{struct} Planar zigzag chains have 2 C atoms per cell
forming apex angles of either $120^\circ$ or 60$^\circ$. Number of
atoms per cell is 4 and 3 for the dumbbell and equilateral
triangular structures, respectively. We also performed
\emph{ab-initio} calculations for finite C-LC having 5, 6, and 20
atoms in a supercell with an axial vacuum distance of 8 \AA\,
between the chains. Optimized structures are found to be stable.

\bibitem{distortion} We note that \textit{ab-initio} Hartree-Fock
(HF) calculations predict $\delta$ in dimerized C chains to be
almost 10 times larger, and the effective potential to be 60 times
deeper as compared to LDA results \cite{jones,abdur02}. However,
from studies on the bond alternation (second order Jahn-Teller
effect) in carbon ring structures it is known that while LDA and
GGA somewhat underestimates the effect, HF method tends to
overestimate it \cite{jones,torelli00}.
%Another
%interesting point to note is that although Si and C have similar
%valency, the $\sigma^\ast$-band dipping the Fermi level
%delocalizes the double band structure, and prevents the Si linear
%chain from undergoing Peierls distortion.

% CNT force constant
\bibitem{portal99} D. Sanchez-Portal \emph{et al}, Phys. Rev. B \textbf{59}, 12678 (1999).

\bibitem{cell} We considered a grid structure with $N=7$, $M=7$.


% C(4N+2) rings
\bibitem{gulseren01} O. G{\"u}lseren \emph{et al}, Phys. Rev. Lett. {\bf 87}, 116802 (2001).


\bibitem{lang}N. D. Lang, and Ph. Avouris, Phys. Rev. Lett. \textbf{81}, 3515 (1998); \textbf{84}, 358 (2000).
\bibitem{hatem} B. Larade \emph{et al}, Phys. Rev. B \textbf{64}, 075420 (2001).
\bibitem{altschuler} B. L. Altschuler and A. G. Aronov  in \textit{ Electron-electron interaction in disordered
systems}, Eds. A. L. Efros and M. Polak, Elsevier Amsterdam
(1985).



\end{thebibliography}
\end{document}